\documentclass[aps,preprint]{revtex4}

\usepackage[dvips]{graphicx}

\usepackage{color}

\usepackage[centerlast]{caption}

\usepackage{setspace}

\begin{document}

\date{\today}

\title{A review of progress in the physics of open quantum systems: 
theory and experiment}

\author{I. Rotter$^{1}$\footnote{email: rotter@pks.mpg.de}
and J. P. Bird$^{2}$\footnote{email: jbird@buffalo.edu}}

\address{$^1$Max Planck Institute for the Physics of Complex Systems,
  D-01187 Dresden, Germany,} 
  
\address{$^2$Department of Electrical Engineering, University at
  Buffalo, the State University of New York, Buffalo, NY 14260, USA}

\begin{abstract}
This Report on Progress explores recent advances in our
theoretical and experimental understanding of the physics of open
quantum systems (OQSs).
The study of such systems represents a core
problem in modern physics that has evolved to assume an unprecedented 
interdisciplinary character. 
OQSs consist of some localized, microscopic, region
that is coupled to an external environment by means of an appropriate
interaction. Examples of such systems may be found in numerous
areas of physics, including atomic and nuclear physics,
photonics, biophysics, and mesoscopic physics. It is the
latter area that provides the main focus of
this review, an emphasis that is driven by the capacity that exists to
subject mesoscopic devices to unprecedented control. 
We thus provide a detailed discussion of the behavior of
mesoscopic devices (and other OQSs) in terms of the 
projection-operator formalism, according to which
the system under study is considered to be comprised of a localized
region ($Q$), embedded into a well-defined environment ($P$) of scattering
wavefunctions (with $Q+P=1$). The $Q$ subspace must be treated using the
concepts of non-Hermitian physics, and of particular interest here is:
the capacity of the
environment to mediate a coupling between the different states of $Q$;
the role played by the presence of exceptional points (EPs) in the
spectra of OQSs; the influence of EPs on the rigidity of the wavefunction
phases, and; the ability of EPs to initiate a dynamical phase transition
(DPT). EPs are singular points in the continuum, at which two
resonance states coalesce, that is where they exhibit a non-avoided crossing. 
DPTs occur when the quantum dynamics of the open system causes transitions
between non-analytically connected states, as a
function of some external control parameter. 
Much like conventional
phase transitions, the behavior of the system on one side of
the DPT does not serve as a reliable indicator of that on the
other. In addition to discussing experiments on mesoscopic quantum
point contacts that provide evidence of the environmentally-mediated
coupling of quantum states, we also review manifestations of DPTs
in mesoscopic devices and other systems. These experiments include
observations of resonance-trapping behavior in microwave cavities and
open quantum dots, phase lapses in tunneling through single-electron
transistors, and spin swapping in atomic ensembles. Other possible
manifestations of this phenomenon are presented, including
various superradiant phenomena in low-dimensional
semiconductors. From these discussions a generic picture of
OQSs emerges in which the environmentally-mediated coupling between
different quantum states plays a critical role in governing the system
behavior. The ability to control or manipulate this interaction may
even lead to new applications in photonics and electronics.

\end{abstract}

\pacs{\bf }

\maketitle

\newpage

\section*{\Large Introduction}

\label{intr}

A core problem in modern physics,
and one which has evolved to assume
an unprecedented interdisciplinary character, is related to the study
of open quantum systems   (OQSs). In the broadest sense, these systems
may be defined as consisting of some localized, microscopic, region 
that is coupled to an external environment by means of an appropriate 
interaction. In discussions of the crossover from quantum mechanics 
to classical physics, the role of such an environment is often invoked 
to describe the influence of a classical measuring apparatus. 
  Even if one is to remove such an apparatus, however, the description
  of most open systems may nonetheless be reduced to one in which the
  properties of some microscopic region are influenced by its coupling
  to its own, ``natural'', environment. In contrast to the
  aforementioned measurement problem, the influence of this
  environment can never be deleted but exists at all times,
  independent of any observer. In essence, this environment functions
  as an ``intrinsic'' (natural) measuring apparatus, introducing 
a coupling between different states of the open system. Two very-different 
cases in which this situation applies
are provided by the decay of unstable states in nuclei,
  and the transport of electrons through mesoscopic quantum dots. In
  the former case, the environment consists of a continuum of
  scattering wavefunctions outside of the nucleus, which may mediate
  the escape of either a neutron or a proton from that structure
  \cite{top}. Similarly, in discussions of quantum-dot transport, the
  discrete quantized states of an isolated cavity develop a broadening
  when the cavity is coupled to macroscopic reservoirs through the
  addition of appropriate leads. In this latter case, the
  environment is provided by the states of these reservoirs, and the
  broadening of the discrete cavity levels is strongly dependent upon
  the mutual overlap with these states, as mediated through the leads  
\cite{ferry review1}.

There are numerous examples of OQSs, from a broad 
spectrum of disciplines within physics. While obvious examples 
include nuclei, atoms, and molecules, 
a much broader range of systems is provided by lasers and
optically-active media, biomolecules and molecular networks,
nanophotonic structures, and mesoscopic electronic devices.
It is the latter systems (namely
mesoscopic devices) that we focus on in this
review, motivated by two of their important features. The first of these is
their capacity to exhibit a variety of rich
phenomena arising from their environmental interaction, while the
second is the ability to subject them to sophisticated external control.

Quite generally, mesoscopic devices are small
  metallic or semiconducting structures, in which carrier motion is
  constrained on a spatial scale that is comparable to, or even
  smaller than, the fundamental length scales associated with
  transport. These length scales include the elastic mean free path,
  the phase-coherence length, and the inelastic scattering length, all
  of which may exceed the size of sub-micron scale devices at
  sufficiently low temperatures \cite{mesoscopic}. In terms of their
  general structure, these devices consist of some central scattering
  region, in the form of a quasi-one dimensional wire or a
  quasi-zero-dimensional quantum dot (QD), whose carriers may be injected
  into and extracted from in order to allow for transport. Such
  transport is achieved by connecting the central scattering region of
  the device to macroscopic charge reservoirs, by means of appropriate
  {\it lead} structures.  When compared with systems such as
  nuclei, the great advantage that mesoscopic devices offer
    is the capacity to externally {\it control} the key parameters of the
    system. These include the system size, and thus the nature of
    the quantum states involved in transport, the strength of the
    coupling between the system and its environment, and the energy of
    the carriers involved in transport. This control makes these
    structures ideally suited to the study of forefront issues
    associated with OQSs. We demonstrate this here for few-electron
QDs \cite{mesoscopic,burosa07}, which are the
solid-state analog of scattering billiards, and for
quasi-one-dimensional quantum point contacts (QPCs), in which the
presence of a strong environmental coupling may modify the very 
nature of the quantum states responsible for transport \cite{rejec}.

\begin{figure}[ht]
\begin{center}
\includegraphics{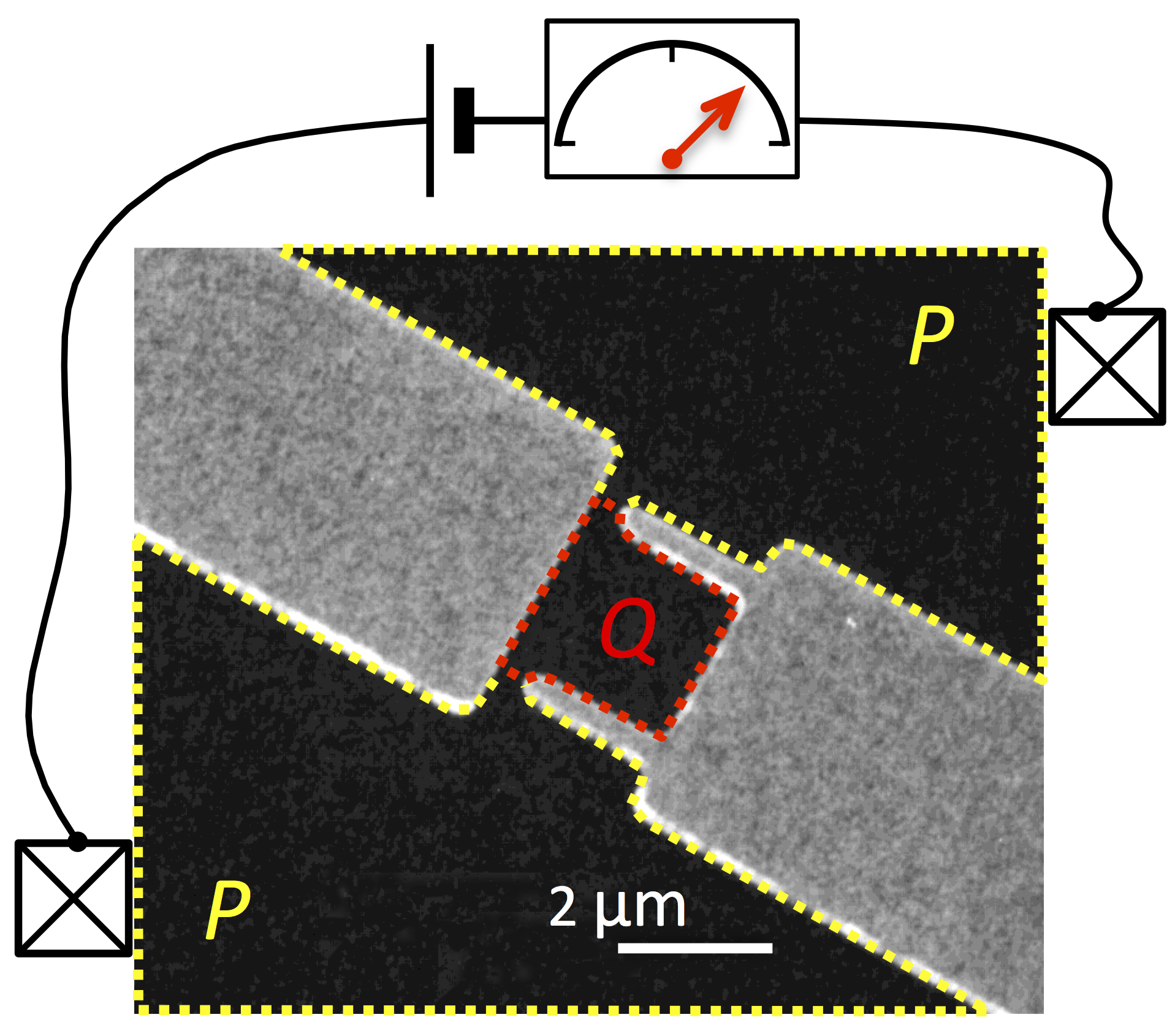}
\end{center}
\caption{\footnotesize
Electron micrograph showing an example of a mesoscopic device, namely
a gated quantum dot. The gray regions in the figure correspond to a
pair of metal gates, which are formed on top of a semiconductor
substrate (the dark areas). Application of a suitable voltage to the
gates creates depletion fields that remove electrons from underneath
the gates, forming a square-shaped cavity approximately 2 $\mu$m in
size. To illustrate the connection to the non-Hermitian concept, we
have also indicated in the figure how this mesoscopic device may
essentially be viewed as consisting of a localized quantum system
(denoted here as the region `$Q$') that is embedded within an
environment of scattering states (denoted as the regions `$P$'). The
coupling between regions $Q$ and $P$ is regulated by means of the two
QPC openings in the cavity, which correspond to the
``leads'' referred to in the main text. In this figure, we have also
indicated the presence of a classical measuring apparatus that is {\it
  outside} of the $P+Q$ space. While this measurement apparatus
provides an ``environment'' that can be removed from the system, the
influence of the environment $P$, in contrast, can {\it never} be
deleted.}
\label{OQS}
\end{figure}

In this {\it Report on Progress}, we focus on providing a detailed
discussion of the behavior of OQSs in terms of the projection operator 
formalism of quantum mechanics. In this approach,
as indicated in Fig. \ref{OQS}, we consider the system under study to
be comprised of a localized region ($Q$) that is embedded in a
well-defined environment ($P$) of scattering wavefunctions. 
The complete function space satisfies the requirement 
$Q+P=1$, and is described in terms of a Hermitian Hamiltonian.
The Hamiltonian describing the subspace of interest ($Q$) is non-Hermitian, however,
and has complex eigenvalues that define both the energy and broadening
of its states \cite{top}. These eigenstates are coherently coupled to each other,
via an interaction that is mediated by the environment ($P$).
This projection operator formalism 
differs from that in which
complex scaling is used to analyze non-Hermitian systems, an approach that
has been summarized in Ref. \cite{moisbook} and that we will
not discuss. We also note that the role of the environment considered here
is very different to that which is often invoked in
discussions of decoherence, arising 
from the uncontrolled interaction of a quantum system with its
environment (such as an electron or phonon bath).
This latter problem has been treated in other works
(see, for example, the reviews of \cite{imry, seligman}), where the
feedback from the environment onto the quantum system is not
explicitly considered. This is in contrast with the
situation here, where we are concerned, above all, with understanding
the influence of the natural environment ($P$) on the system ($Q$).
This environmental influence remains
even when sources of decoherence have been suppressed (by lowering the
temperature, for example). As such, the approach
that we outline is expected to be useful in
describing  the behavior of {\it small} systems, in which
the number of particles active in transport is small, and in which the
number of quantum states they occupy is similarly restricted in
number. Under such conditions we will see that, far from acting as 
a source of decoherence, the coupling to the environment can in fact mediate an
extremely robust and coherent interaction.

The objective of this review is to focus on critical implications of non-Hermiticity, 
namely the presence of {\it exceptional points} (EPs) in the spectrum 
of OQSs, their influence on the {\it rigidity} of the wavefunction 
phases, and their ability to initiate a {\it dynamical phase
transition} (DPT). As we discuss in further detail below, 
these features can be observed because of our ability to vary the 
strength of the environmental interaction in OQSs, by means of some 
suitable control parameter. EPs are singular points in the continuum, 
at which two resonance states coalesce, that is where they exhibit 
a non-avoided {\it crossing}. This behavior should be contrasted with the 
well-known {\it avoided} crossings, exhibited by Hermitian systems
at so-called {\it diabolic points}. Due to the coalescence of the two
resonance states at an EP, the time evolution of the system
becomes undefined in its neighborhood.
At the same time, the EPs are branch points in the
complex plane and are responsible for the generation of double poles 
of the $S$ matrix, with associated implications for transmission.

An important quantity that indicates the range of influence of an EP
is the phase rigidity of its two eigenfunctions. In isolated
(Hermitian) systems the phases of the eigenfunctions are said to be
rigid, a statement that expresses the well-known orthogonality
of these functions. In open (non-Hermitian)
systems, however, the presence of the environmentally-mediated
coupling between the
eigenstates means that they are no longer orthogonal, but rather
exhibit the property of {\it biorthogonality}. This allows the
rigidity of the eigenfunction phases to be reduced, particularly
near an EP where the rigidity actually vanishes. This
surprising behavior expresses the very strong influence of
 the environment on the localized system at an EP.  

While the physical significance of EPs
for the dynamics of OQSs has only recently begun to be studied
(see, for example, the review  \cite{top}), they are
nonetheless understood to be
intimately connected to the observation of DPTs in different open
systems. The term DPT has been coined to refer to a phenomenon in
which the quantum dynamics of the open system undergoes a phase
transition between non-analytically connected states as a
function of some suitable control parameter.  Here, time-dependent
approaches to the description of the problem fail \cite{fdp1}. 
Instead, much like a conventional
phase transition in, for example, magnetism or superconductivity, the
physical behavior of the system on one side
of the transition does not serve as a reliable indicator of that on
the other. The connection of DPTs to the EPs is provided by
the phenomenon of {\it width bifurcation}, which results in a
spectroscopic redistribution in the system into long-lived states
with narrow linewidths and a 
much smaller number of short-lived, strongly-broadened, ones. An
important objective of this review is to connect the theoretical
concepts of EPs and DPTs to the physical behavior manifested in
experimental investigations of OQSs.

One of the earliest works to
explicitly invoke the notion of a DPT involved studies of spin
swapping in atomic systems \cite{past1}. In these experiments, the
authors studied Rabi oscillations due to spin flips in the
$^{13}$C--$^{1}$H system, and showed a transition to strongly damped
motion by increasing the coupling to an environment formed by a spin
bath. Although not explicitly discussed as such, evidence of DPTs is
apparent, also, in earlier work on resonance trapping in
microwave cavities \cite {persson}, and on
einselection \cite {bird,akis,ferry} in mesoscopic quantum dots.
 In both of these  systems, the focus was on understanding
the manner in which the states of a quantized cavity are affected
by coupling them to an external environment. The common
  phenomenon revealed in both cases was of width bifurcation,
with certain eigenstates actually becoming narrower when the 
environmental coupling was increased over a specific range.
The accompanying short-lived eigenstates appear as a background with
which the long-lived ones interfere, a behavior that has been 
observed experimentally in microwave cavities \cite{persson}.
Ultimately, the long-lived states may even become discrete,
forming so-called {\it bound states in the continuum} \cite{bicrosa}.
With the renewed interest in such problems, it is actually interesting
to note that the appearance of long-lived 
``merkw\"urdigen'' (remarkable) states is actually a problem that dates back 
to the earliest days of quantum mechanics \cite{wigner}!

Elsewhere, evidence of a DPT has also been provided in work where a 
multi-level quantum dot was embedded into one of the arms of an
Aharonov-Bohm interferometer, allowing the evolution of the
transmission phase to be monitored across a sequence of 
resonance states \cite{yacoby,schuster,avinun}. These
experiments revealed the presence of unexpected regularity in the
measured scattering phases (so-called ``phase lapses"), when the number of states
occupied by electrons in the dot was sufficiently large.
While this behavior could not
be fully explained within approaches based upon Hermitian quantum
theory \cite{laps2,oreg}, it has recently been established that
the phase lapses can be attributed to the non-Hermitian character of
this mesoscopic system, and to a DPT that occurs as the number of
electrons in the dot is varied \cite{laps3}.
The observed regularity arises from the overlap of the many
long-lived states with the short-lived one, all of which are formed due to
the DPT. 
More recently, studies of transport in ballistic quantum wires may have also
revealed a DPT, involving the formation of a protected channel for
conduction under strongly nonequilibrium conditions. In these
experiments the environment is
essentially provided by the phonon system, whose influence is
controlled by means of the applied source bias and by the 
strong quantum confinement of
the carriers induced within the wire \cite{bird14}. A well-known
phenomenon from optics that may also involve a DPT is that of Dicke
superradiance \cite{dicke}. This refers to the effect in which an
  ensemble of excited atoms within a cavity does not emit radiation
  randomly, but rather does so in a correlated manner when the atoms
  experience the same radiation field. Recently, superradiance
  has been demonstrated for solid-state systems, namely ensembles of
  self-assembled quantum dots \cite{scheibner} and a dense
  semiconductor electron-hole plasma \cite{Noe}.

In this review, we will use the various experiments described above to
connect the behavior exhibited by OQSs to the key concepts
of non-Hermiticity, EPs, and DPTs. We will also introduce the important concept
of wavefunction phase rigidity, and the significance of
time in open systems,  in which we are unable to describe a DPT by means
  of time-dependent approaches. We emphasize
  again that the focus of our review will be on discussions of the
  properties of open systems consisting of a small number of
  particles. As such, we explore the behavior in a very different
  limit to that relevant, for example, in heavy nuclei. These contain a
much larger number of particles, as well as many
  closely-neighboring states, and are adequately described by the
  concepts of random-matrix theory \cite{brody}. In fact, with the
  exception of a few specific examples, we will not address phenomena
  arising in nuclear systems, preferring to note instead
  that this topic was recently excellently served by an associated
  review in this journal \cite{auerbach}.  Due to the continued
  importance of mesoscopic systems for the investigation of 
fundamental quantum phenomena, we believe that the focus of our review 
will prove to be a particularly useful one.

The remainder of this paper is organized as follows. In the next section, 
we introduce the concept of small quantum systems that are coupled to an
environment with which they interact. We begin by reviewing some of the
well known properties of isolated (Hermitian) quantum systems, emphasizing
concepts such as the real nature of their eigenvalues and the rigidity
(orthogonality) of their eigenfunctions. Following this, we next introduce the
projection-operator formalism for OQSs, in which the total system is
described in terms of two coupled subspaces ($Q$ \& $P$).
The complex character of the eigenvalues and eigenfunctions of 
the non-Hermitian operator of the $Q$ subspace is
described, and its physical implications are discussed.
We introduce also the concept of gain and loss in OQSs. In the last
part of this section, some physical examples of the 
environmentally-mediated coupling of
quantum states are presented, focusing on their manifestations in 
mesoscopic structures. In Section III, we discuss
the importance of EPs to the behavior of OQSs. We formulate this
discussion by first of all focusing on the interaction of two resonance
states near an EP, following which we study the influence of an EP 
on the eigenvalues and eigenfunctions of the non-Hermitian Hamiltonian.
In Section III\,D we discuss the implications of EPs for the 
$S$-matrix of the system, a problem with important implications for
the analysis of transmission. In Section IV, we consider the role of EPs
in giving rise to DPTs in open systems subjected to some form of external control.
In the presence of this control, DPTs are found to occur when the
range of influence of several different EPs develops a sufficiently strong overlap.
The connection of DPTs to the notion of width bifurcation is
emphasized, and several experimental demonstrations of such phase transitions
are presented. These include demonstrations of resonance trapping in
microwave cavities, studies of phase lapses in tunneling through quantum
dots, and measurements of spin swapping in atomic systems. Emphasizing
our belief that DPTs are, in fact, an inherent feature of OQSs in general,
in Section V we discuss further examples of physical problems in
which DPTs may occur. In Section VI, we conclude this review
by summarizing its main points and by identifying important issues
for further study. Two appendices to the paper are provided, in the first of which
(Appendix A) we make a few additional remarks on the non-Hermitian
Hamitonian (with and without using perturbation theory), and give
expressions for the coupling matrix elements (the partial width
amplitudes) between the system and the environment. Appendix B
provides a list of principal symbols and acronyms used in the paper.

\newpage

\section*{\Large Conclusions}

\label{concl}

In this Report on Progress, we have reviewed the key physical
attributes of OQSs in the regime of closely-neighboring states
and have described a number of different
experimental systems in which they are manifested. At the
most basic level, these attributes may be understood as arising from
the fact that the Hamiltonians needed to describe the dynamics of open
systems are non-Hermitian, in marked contrast to their Hermitian
counterparts that so accurately describe the behavior of isolated
quantum systems. In the non-Hermitian formalism, the open system is
viewed as being comprised of a localized region ($Q$), embedded in a 
well-defined environment ($P$) of scattering wavefunctions, and the
total system ($Q+P=1$) is Hermitian. The coupling between the two 
subspaces is found to mediate an effective interaction among the 
original states of the localized system. This
{\it natural} ({\it intrinsic}) environment does not serve as a source 
of decoherence (the role of which is not considered here)
and its influence  can never be deleted from
the system. Instead, it  gives rise to complex eigenvalues, 
whose real component defines the energy of the state
and whose imaginary part describes an effective level broadening.
The eigenfunctions also exhibit an important difference with those of
Hermitian systems, in that they are no longer orthogonal to one
another but are instead biorthogonal. Physically, the biorthogonality
expresses the fact that there is an environmentally-induced coupling 
between the different states of the system. This should be contrasted 
with the case of Hermitian quantum mechanics, which yields orthogonal
stationary states that require a perturbation of the system in order to be
coupled. The description of open systems in terms of this approach
is found to be well suited to the solution of a broad
range of problems, providing proper care is taken to accurately
identify the subspaces $Q$ and $P$. Under such conditions, this approach is
expected to provide a good description of small systems,
in which the number of relevant quantum states, as well as the number of
channels, is restricted. In larger systems that do not conform to these
constraints, other approaches, such as random-matrix theory, should be
more effective. Such approaches have been extensively treated in the
literature and have therefore not been considered here. 

We have seen in this review how the eigenstates of the non-Hermitian
Hamiltonian exhibit fundamentally new behavior not encountered in Hermitian
physics. The first example is provided by the appearance of singular
points, referred to as exceptional points (EPs). Arising when an
appropriate system parameter (such as the coupling strength between
the $Q$ and $P$ subspaces) is varied, the EP involves the coalescence 
of two eigenvalues of the non-Hermitian operator and (up to a phase) 
their corresponding eigenfunctions. Although a point of measure zero, 
the EP exerts a strong
influence on the spectral properties of the system, extending over a
 wider region than the singular point at which the two eigenvalues coalesce.
An important property that indicates the
 range of influence of the EP is the phase rigidity of the
 eigenfunctions. Essentially, the phase rigidity is a measure of the
 biorthogonality of different eigenfunctions, and, therefore, of the
 strength of the environmentally-induced interaction between the
 different states. In the region of the EP the phase rigidity is
 significantly reduced, expressing the fact that this interaction 
via the environment is very strong at the EP. As a result, the 
wavefunctions of these states are no longer
 orthogonal with one another but are equal,  up to a phase. Over 
the range where the phase rigidity is reduced, the two eigenstates 
undergo the phenomenon of width bifurcation, with one state becoming
strongly broadened while the other becomes more stable as it decouples 
from the environment. Under this condition, transport through
the system occurs with high efficiency. Simultaneous with this, the
eigenvalues may cross in energy. This is in complete contrast to the
behavior of Hermitian systems, which may only ever exhibit an avoided
level crossing when two of their eigenstates are close in energy.

The phenomenon of width bifurcation, in which the width of one state
increases with increasing environmental coupling while that of the
other decreases, is connected to a violation 
of the Fermi golden rule. This rule suggests that increased coupling 
should always lead to stronger broadening, in contrast to the behavior
of the long-lived state produced by width bifurcation. 

The other important feature of OQSs that we have discussed is their ability
to undergo a dynamical phase transition (DPT), when the ranges of
influence of different EPs overlap with one another. The DPT involves
a change in the dynamics of the system, which arises when width
bifurcation occurs hierarchically for many states. The simplest
example of this phenomenon occurs when all states of the system are coupled to a
single channel, resulting in the formation of just one strongly-broadened state.
The remaining states are then essentially stabilized, through a
process in which they strongly decouple from the environment. In the
two-channel case relevant for transport, essentially the same picture 
holds although now two of the original states develop the strong
broadening, leaving all but these two states stabilized. Much like
conventional phase transitions in thermodynamics, the details of the
dynamics on either side of the DPT are not analytically connected to
each other. That is, the behavior on one side of the transition 
does not serve as a reliable indicator of that on the other. Both the
width bifurcation caused by the EPs, and the DPT that these
give rise to, result from the fact that the coupling strength to the
environment has an imaginary component.

Experimental evidence in support of these various concepts has been
provided throughout the course of this review, with a particular
emphasis having been placed on their manifestations in mesoscopic
systems. At the heart of the non-Hermitian scheme is the idea of
environmentally-mediated coupling, which yields an effective
interaction among the  states of the quantum system. We have
seen how this coupling is manifested directly in experiments involving
non-locally coupled quantum point contacts (QPCs), in which an intervening
continuum serves as the $P$ subsystem and mediates a coupling between
remote pairs of QPCs. We have described how this coupling gives rise
to novel Fano-resonance phenomenology, and have also observed that the
environmentally-mediated interaction may provide a stronger coupling
between quantum states than a simple tunneling overlap. This may be
attributed to the fact that the environmental interaction is supported
by a large number of states that comprise a continuum, resulting in a
much stronger effect than would be expected to arise from the direct
interaction between the different states of $Q$.

Several experimental demonstrations of DPTs in physical systems were
also presented. These examples included: observations of
resonance-trapping and its related phenomena, in open microwave
cavities and QDs; a crossover from mesoscopic to universal behavior in
the phase lapses found in single-electron tunneling through
Coulomb-blockaded QDs; observations of spin-swapping dynamics in
ensembles of interacting spins,
and; energy-transfer processes in photosynthesis. 
Similar phenomena are observed also in optics, most notably in the form of
the long-studied problem of Dicke superradiance, 
as well as more recent investigations of systems with gain and loss. Here, however, a
better understanding of the equivalence between the optical-wave
and Schr\"odinger equations is needed before convincing conclusions can be drawn.
In addition, we have also 
highlighted several other examples of complex physical phenomena 
that may well be related to the occurrence of a DPT. These include 
superradiance in QDs, and the observation of
protected-subband formation in the nonequilibrium transport of hot
electrons through QPCs. These observations by no means represent an
exhaustive list, but rather suggest instead that DPTs are a common
property of OQSs in general.

Having summarized the general status of this field, we can now
identify a number of important issues that should be resolved in
future studies. While the different experiments described in this 
review provide several hints that DPTs do indeed occur in OQSs, the
presence of the distinct time scales that are expected to accompany
this phenomenon has yet to be directly demonstrated. According to the description
of width bifurcation given above, the environmentally-induced
interaction in the vicinity of an EP is expected to result in the
formation of eigenstates with very different lifetimes; the first of
these is very short and associated with the strongly-broadened state,
while the other is much longer and related to the stabilized
state. Experiments that can probe the existence of these different
time scales are therefore desired. A problem in attempting to design
such experiments, however, comes from the fact that this problem
must first be clearly formulated from a theoretical perspective.
A simpler task may be to provide a demonstration of the width
bifurcation that should serve as the signature of the different time
scales, but, to the best of our knowledge, this has not yet been done
experimentally, in either mesoscopic or any other physical systems.

Time-resolved investigations of OQSs may also be useful in exploring
issues associated with irreversibility in the vicinity of the DPT. The
 issues here can be discussed by referring to the features of
Fig. 9, in which we show the system properties and their
variation in the vicinity of two EPs. As either EP is approached from
the left or the right of the figure, the width bifurcation begins at the EP and
becomes maximal once $d=0$. At this point the phase rigidity returns
to unity, indicating that the two eigenfunctions are
orthogonal. Physically, this can be understood to result from the
fact that the two states are now associated with very different time
scales, and so no longer interact with one another. While the results of
Fig. 9 suggest that, by passing continuously through $d=0$ and
the point of maximum bifurcation, it should be possible to return to
two distinct states, in practice this is not possible. In essence, one
of the original states has now been ``lost'' to the environment, and
the remaining stable state is therefore unable to interact directly with
it. This is completely consistent with the notion of a DPT; the
spectroscopic properties of the system on one side of the transition 
are not analytically connected to those on the other, and nonlinear
terms in the Schr\"odinger equation play an important role in this 
transition.

Another important property that is of interest in characterizing the
properties of OQSs is the phase rigidity of their eigenfunctions. The
value of this parameter becomes vanishingly small near an EP, and it
is useful also for establishing the range of influence of such
points. While the rigidity itself can probably not be measured 
directly in a physical system, we have seen here that it is related to 
a high efficiency of transport (recall Fig. 13). 
While this connection may not have been widely
appreciated to date, it implies that measurements of the conductance
of mesoscopic systems can provide a flexible probe of this important
system property. Moreover, the enhancement of the conductance
  around a DPT is, by itself, of interest for possible applications. 

Throughout this review, we have provided a number of different
examples of DPTs, occurring in a variety of physical systems. From these
examples we suggest that the DPT should be a very general characteristic of
open systems, with broader consequences than have been appreciated
to date. Indeed, the stability properties of classical
systems are also governed by non-Hermitian degeneracies (EPs),
as has been nicely shown in \cite{kirillov}.
Our hope therefore is that this review can stimulate interest in the
study of this still relatively unexplored problem, and ultimately lead to a
deeper understanding of its implications. \\ 

{\it Acknowledgment}: 
We acknowledge the support of the Max Planck Institute for the Physics of 
Complex Systems, which enabled JPB to visit Dresden twice during the 
preparation of this manuscript. JPB is supported by the US Department of 
Energy, Office of Basic Energy Sciences, Division of Materials Sciences 
and Engineering (award DE-FG02-04ER46180).\\

{\it Note added in proof}. Subsequent to the acceptance of our report
we became aware of recently published work by Ruderman et al. 
\cite{ruderman},
demonstrating the possibility of a quantum dynamical phase transition
in molecular chemical bond formation and dissociation. The occurrence
of this transition was connected to the non-Hermitian nature of the
Hamiltonian in the molecular system, consistent with the dicussions of
dynamical phase transitions given here.

\end{document}